\documentclass[aps,prd,preprint,amsmath,amssymb]{revtex4}
\usepackage{graphicx}
\usepackage{dcolumn}
\usepackage{bm}

\newcommand{\be}{\begin{equation*}}
\newcommand{\ee}{\end{equation*}}
\newcommand{\bea}{\begin{eqnarray}}
\newcommand{\eea}{\end{eqnarray}}
\newcommand{\bean}{\begin{eqnarray*}}
\newcommand{\eean}{\end{eqnarray*}}

\begin{document}

\title{Prediction of super-heavy $N^*$ and $\Lambda^*$ resonances with hidden beauty}

\author{Jia-Jun Wu$^{1}$, Lu Zhao$^{1}$ and B.~S.~Zou$^{1,2}$ \\
$^1$ Institute of High Energy Physics, CAS, P.O.Box 918(4), Beijing 100049, China\\
$^2$ Theoretical Physics Center for Science Facilities, CAS,
Beijing 100049, China}

\date{June 26, 2011}

\begin{abstract}
The meson-baryon coupled channel unitary approach with the local
hidden gauge formalism is extended to the hidden beauty sector. A
few narrow $N^*$ and $\Lambda^*$ resonances around 11~GeV are
predicted as dynamically generated states from the interactions of
heavy beauty mesons and baryons. Production cross sections of these
predicted resonances in $pp$ and $ep$ collisions are estimated as a
guide for the possible experimental search at relevant facilities.
\end{abstract}


\maketitle

\section{Introduction}
\label{s1}

In conventional quark models, all established baryons are ascribed
into simple 3-quark (qqq) configurations~\cite{PDG}. The excited
baryon states are described as excitation of individual constituent
quarks, similar to the cases for atomic and nuclear excitations.
However, unlike atomic and nuclear excitations, the typical hadronic
excitation energies are comparable with constituent quark masses.
Hence to drag out a $q\bar q$ pair from gluon field could be a new
excitation mechanism besides the conventional orbital excitation of
original constituent quarks. Some baryon resonances are proposed to
be meson-baryon dynamically generated states~\cite{Weise,or,Oset,
meiss,Inoue, lutz,Hyodopk} or states with large ($qqqq\bar q$)
components~\cite{Riska,Liubc,Zou10}. A difficulty to pin down the
nature of these baryon resonances is that the predicted states from
various models are around the same energy region and there are
always some adjustable ingredients in each model to fit the
experimental data. A typical example is $N^*(1535)$ which has large
couplings to the strangeness. In the 3-quark (qqq) configurations,
it is described as the orbital angular momentum $L=1$ excitation of
a quark. But phenomenological studies suggest that it may be a
quasi-bound state of $K\Sigma$ system~\cite{siegel,inoue,Nievesar},
or a hidden strangeness 5-quark state~\cite{Liubc,Geng:2008cv}. In
order to clearly demonstrate the new excitation mechanism with some
of its corresponding states, in Ref.\cite{charm}, the meson-baryon
coupled channel unitary approach with the local hidden gauge
formalism was performed for the hidden charm sector and several
narrow $N^*$ and $\Lambda^*$ resonances with hidden charm were
predicted to exist. If found experimentally, these resonances
definitely could not be described as three constituent quark states.
Here, we extend the study to the hidden beauty sector. Some
super-heavy $N^*$ and $\Lambda^*$ resonances with hidden beauty are
predicted to exist, with mass around 11 GeV and width smaller than
10 MeV. If these resonances would be experimentally confirmed, they
should be part of the heaviest super-heavy island of $N^*$ and
$\Lambda^*$ state. As a guild to the future experimental search for
these new predicted states, their production cross sections in $pp$
and $ep$ collisions are estimated.

In the next section, we present the formalism and ingredients for
the study of interactions between heavy beauty meson and baryon with
the Valencia approach, and give some detailed discussion on the
intermediate meson-baryon loop G functions. In section~\ref{s3}, our
numerical results for the masses and widths of the predicted
super-heavy $N^*$ and $\Lambda^*$ states are given, followed by a
discussion. In section~\ref{s4}, effects of momentum dependent terms
in the effective potential are investigated. In section~\ref{s5},
the calculation about production of these predicted states from $pp$
and $ep$ collisions is presented. Finally, a short summary is given
in the last section.

\section{Formalism for Meson-Baryon Interaction}
\label{s1}

\begin{figure}[htbp] \vspace{-0.cm}
\begin{center}
\includegraphics[width=0.7\columnwidth]{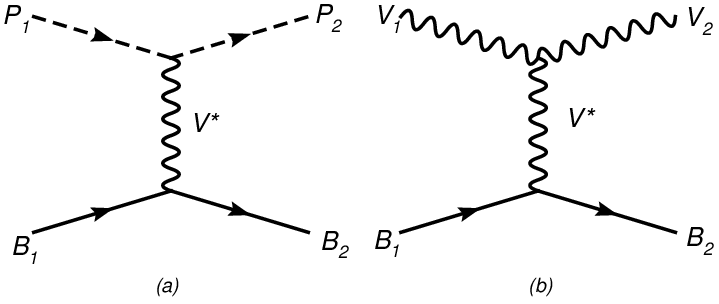}
\caption{Feynman diagrams for the pseudoscalar-baryon (a) or
vector-baryon (b) interaction via the exchange of a vector meson
($P_{1}$, $P_{2}$ are $B^0$, $B^+$ or $B^{0}_{s}$, and $V_{1}$,
$V_{2}$ are $B^{0*}$, $B^{+*}$ or $B^{0*}_{s}$, and $B_{1}$,
$B_{2}$ are $\Sigma_{b}$, $\Lambda_{b}$, $\Xi_{b}$, $\Xi'_{b}$ or
$\Omega_{b}$, and $V^{*}$ is $\rho$, $K^{*}$, $\phi$ or
$\omega$).} \label{fe}
\end{center}
\end{figure}

We follow the recent work of Ref.~\cite{charm} on the interactions
between charmed mesons and baryons, and replace charm quark by
beauty quark. The $PB\to PB$ and $VB\to VB$ interactions by
exchanging a vector meson are considered, as shown by the Feynman
diagrams in Fig.~\ref{fe}.

The effective Lagrangians for the interactions involved
are~\cite{ramos}:
\begin{eqnarray}
{\cal L}_{VVV}&=&ig\langle V^\mu[V^{\nu},\partial_\mu V_{\nu}]\rangle\nonumber\\
{\cal L}_{PPV}&=&-ig\langle V^\mu[P,\partial_\mu P]\rangle\nonumber\\
{\cal L}_{BBV}&=&g (\langle\bar{B}\gamma_\mu
[V^\mu,B]\rangle+\langle\bar{B}\gamma_\mu B\rangle\langle
V^\mu\rangle)\ \label{eq:lag}
\end{eqnarray}
where $P$ and $V$ stand for pseudoscalar and vector mesons of the
16-plet of SU(4), respectively.

Using the same approach of Ref.\cite{charm}, we keep only the
$\gamma^{0}$ component of Eq.(\ref{eq:lag}), while the three
momentum versus the mass of the meson can be neglected under the low
energy approximation. Similarly, the $q^2/M^2_V$ term in the vector
meson propagator is neglected so that the propagator is
approximately equal to $g^{\mu\nu}/M^{2}_{V}$. Note when we consider
transitions from heavy mesons to light ones later on, we perform the
exact calculation without such approximation. Then with $g=M_V/2f$
the transition potentials corresponding to the diagrams of Fig.
\ref{fe} are given by
\begin{eqnarray}
V_{ab(P_{1}B_{1}\to P_{2}B_{2})}&=&\frac{C_{ab}}{4f^{2}}(E_{P_{1}}+E_{P_{2}})\label{vpbb},\\
V_{ab(V_{1}B_{1}\to
V_{2}B_{2})}&=&\frac{C_{ab}}{4f^{2}}(E_{V_{1}}+E_{V_{2}})\vec{\epsilon}_1\cdot\vec{\epsilon}_{2},\label{vvbb}
\end{eqnarray}
where the $a,b$ stand for different channels of $P_{1}(V_{1})B_{1}$
and $P_{2}(V_{2})B_{2}$, respectively. The $E$ is the energy of
corresponding particle. The $\vec{\epsilon}$ is the polarization
vector of the initial or final vector. And the $\epsilon_{1,2}^{0}$
component is neglected consistently with taking $\vec{p}/M_V\sim 0$,
with $\vec{p}$ the momentum of the vector meson. Here we only change
the charm quark to beauty quark, so the $C_{ab}$ coefficients are
exactly the same as those in Ref.\cite{charm}, so that there are
only two cases, (I, S) = (1/2, 0) and (0, -1), which have attractive
potentials. We list the values of the $C_{ab}$ coefficients for
$PB\to PB$ for these two cases in Table I and Table II,
respectively.

\begin{table}[htbp]
\renewcommand{\arraystretch}{1.2}
\centering \caption{ Coefficients $C_{ab}$ in Eq.
(\protect\ref{vpbb}) for $(I,S)=(1/2, 0)$}\label{zcof}
 \vspace{0.cm}
\begin{tabular}{l|cccccccc}
 & $B \Sigma_{b}$ & $B \Lambda_{b}$ & $\eta_{b} N$  & $\pi N$ & $\eta N$ & $\eta' N$ & $K \Sigma$ & $K \Lambda$\\
 \hline
$B \Sigma_{b}$     & $-1$  &  $ 0$   & $-\sqrt{3/2}$  & $-1/2$  &  $-1/\sqrt{2}$  &   $1/2$   &     $1 $  &  $  0$          \\
$B \Lambda_{b}$    &       &  $ 1$   &  $\sqrt{3/2}$  & $-3/2$  &  $1/\sqrt{2}$   &  $-1/2$   &     $0$   &   $ 1$        \\
\end{tabular}
\end{table}

\begin{table}[htbp]
 \renewcommand{\arraystretch}{1.2}
\centering \caption{ Coefficients $C_{ab}$ in Eq.
(\protect\ref{vpbb}) for $(I,S)=(0,-1)$}\label{mcof} \vspace{0.cm}
\begin{tabular}{l|cccccccccc}
 & $B_{s} \Lambda_{b}$ &  $B \Xi_{b}$ & $B \Xi^{'}_{b}$ & $\eta_{b}\Lambda$ & $\pi \Sigma$     &  $\eta \Lambda$    & $\eta' \Lambda$   & $\bar{K}N$     & K $\Xi$             \\
 \hline
   $B_{s} \Lambda_{b}$ & $0$& $-\sqrt{2}$ & $0$ & $1$                   & $0$                 & $\sqrt{\frac{1}{3}}$  &  $\sqrt{\frac{2}{3}}$    &  $-\sqrt{3}$ & $0$\\
   $B \Xi_{b}$         &    &  $-1$       & $0$ & $\sqrt{\frac{1}{2}}$  & $-\frac{3}{2}$      & $\sqrt{\frac{1}{6}}$  & $-\sqrt{\frac{1}{12}}$   &  $ 0$        & $\sqrt{\frac{3}{2}}$  \\
   $B \Xi^{'}_{b}$     &    &             & $-1$& $-\sqrt{\frac{3}{2}}$ & $\sqrt{\frac{3}{4}}$& $-\sqrt{\frac{1}{2}}$ &  $\frac{1}{2}$           &  $0$         & $\sqrt{\frac{1}{2}}$\\
   $\eta_{b}\Lambda$   &    &             &     &$0$                    &  $0$                & $ 0$                  & $0$                      &   $0$        &  $0$\\
\end{tabular}
\end{table}

With the transition potential, the coupled-channel scattering matrix
can be obtained by solving the coupled-channel Bethe-Salpeter
equation in the on-shell factorization approach of
Refs.\cite{or,meiss}
\begin{eqnarray}
T=[1-VG]^{-1}V \label{Bethe}
\end{eqnarray}
with G being the loop function of a meson (P), or a vector (V), and
a baryon (B). The $\vec{\epsilon}_1\cdot\vec{\epsilon}_2$ factor of
Eq. (\ref{vvbb}) factorizes out also in $T$.

For the G loop function, there are usually two ways to regularize
it. The first one is using dimensional regularization by means of
the formula
\begin{eqnarray}
G&\!=\!&i2M_{B}\int\frac{d^{4}q}{(2\pi)^{4}}\frac{1}{(P\!-\!q)^{2}
\!-\!M^{2}_{B}\!+\!i\varepsilon}\frac{1}{q^{2}\!-\!M^{2}_{P}\!+\!i\varepsilon},\nonumber\\
&=&\frac{2M_{B}}{16\pi^2}\big\{a_{\mu}+\textmd{ln}\frac{M^{2}_{B}}{\mu^{2}}
+\frac{M^{2}_{P}-M^{2}_{B}+s}{2s}\textmd{ln}\frac{M^{2}_{P}}{M^{2}_{B}}\nonumber\\
&&+\frac{\bar{q}}{\sqrt{s}}\big[\textmd{ln}(s-(M^{2}_{B}-M^{2}_{P})+2\bar{q}\sqrt{s})+\textmd{ln}(s+(M^{2}_{B}-M^{2}_{P})+2\bar{q}\sqrt{s})\nonumber\\
&&-\textmd{ln}(-s-(M^{2}_{B}-M^{2}_{P})+2\bar{q}\sqrt{s})-\textmd{ln}(-s+(M^{2}_{B}-M^{2}_{P})+2\bar{q}\sqrt{s})\big]\big\}\
,\label{Gf}
\end{eqnarray}
where $q$ is the four-momentum of the meson, $P$ is the total
four-momentum of the meson and the baryon, $s=P^2$, $\bar q$
denotes the three momentum of the meson or baryon in the center of
mass frame, $\mu$ is a regularization scale, which we put 1000 MeV
here. Changes in the scale are reabsorbed in the subtraction
constant $a_{\mu}$ to make results scale independent.  $a_{\mu}$
is of the order of $-2$, which is the natural value of the
subtraction constant \cite{ollerulf}. When we look for poles in
the second Riemann sheet, we should change $q$ to $-q$ when
$\mathrm{\sqrt{s}}$ is above the threshold in
Eq.(\ref{Gf})~\cite{luisaxial}.

The second way to regularize the $G$ loop function is by putting a
cutoff in the three-momentum:
\begin{eqnarray}
G&=&i2M_{B}\int\frac{d^{4}q}{(2\pi)^{4}}\frac{1}{(P-q)^{2}-M^{2}_{B}+i\varepsilon}\frac{1}{q^{2}-M^{2}_{P}+i\varepsilon}\nonumber\\
&=&\int^{\Lambda}_{0}\frac{\bar q^{2}d\bar
q}{4\pi^{2}}\frac{2M_{B}(\omega_{P}+\omega_{B})}{\omega_{P}\,\omega_{B}\,(s-(\omega_{P}+\omega_{B})^{2}+i\epsilon)}\
,\label{Gf2}
\end{eqnarray}
where $\omega_{P}=\sqrt{\bar q^{2}+M^{2}_{P}}$,
$\omega_{B}=\sqrt{\bar q^{2}+M^{2}_{B}}$, and $\Lambda$ is the
cutoff parameter in the three-momentum of the function loop.

\begin{figure}[htbp] \vspace{-0.cm}
\begin{center}
\includegraphics[width=0.49\columnwidth]{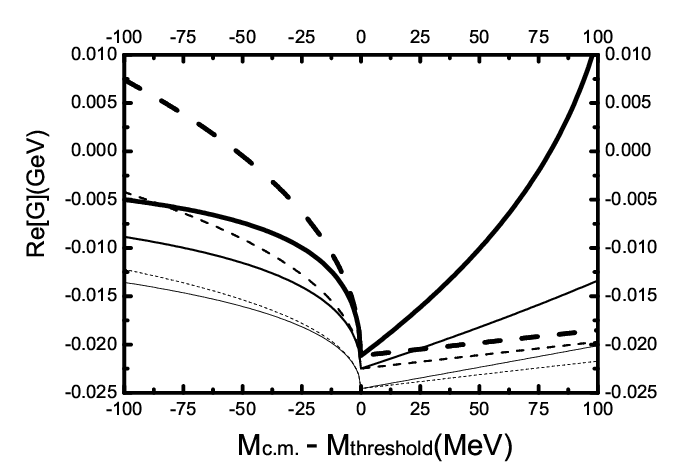}
\includegraphics[width=0.49\columnwidth]{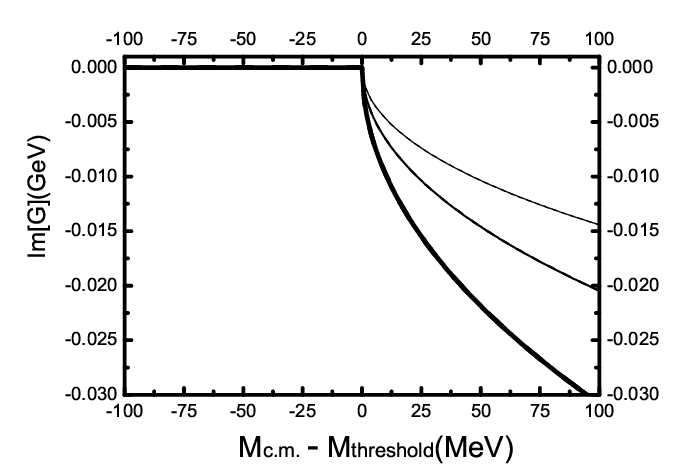}
\caption{The real part (left) and imaginary  part (right) of two G
functions vs the energy difference between the C.M. energy and the
threshold energy. The solid lines are for Eq.(\protect\ref{Gf2}),
and dashed lines are for Eq.(\protect\ref{Gf}). The thickest lines
are for $B\Sigma_b$ channel, the thinnest ones are for $K\Sigma$
channel, and middle ones are for $\bar{D}\Sigma_c$ channel. The
used parameters are listed in the Table.\protect\ref{gfunt} with
$\Lambda=0.8GeV$.} \label{pgf}
\end{center}
\end{figure}

Here we give some detailed discussion on these two types of G
function. Firstly the free parameters are $a_{\mu}$ in Eq.(\ref{Gf})
and $\Lambda$ in Eq.(\ref{Gf2}). The value of $\Lambda$ is around
0.8 GeV, which are within the natural range for effective theories
\cite{meiss}. Then the $a_\mu$ parameter is determined by requiring
that the two G functions from Eq.(\ref{Gf}) and Eq.(\ref{Gf2})take
the same value at threshold. This value also leads to similar shape
near threshold for the two G functions as shown in Fig.\ref{pgf}. In
Fig.\ref{pgf}, the real part and imaginary part of two G functions
vs the energy difference between the center mass energy and the
corresponding threshold for $K\Sigma$, $\bar D\Sigma_c$ and
$B\Sigma_b$ channels are demonstrated. In Table.\ref{gfunt}, the
parameters for different G functions and channels are listed. While
the imaginary parts of two G functions are exactly the same, there
are some differences for the real parts of two G functions and the
differences become bigger for heavier channels. For the same
$\Lambda$ value, the magnitude of $a_\mu$ depends on the threshold
of channels and gets bigger for heavier channels. One point worth
mentioning is that for the $B\Sigma_b$ channel the real part of the
G function given by Eq.(\ref{Gf}) is larger than zero for energies
more than 50 MeV below the threshold as shown in Fig.\ref{pgf}. As
we know, if the interaction is repulsive potential, {\sl i.e.}, the
value of the potential $V$ is positive, there should be no bound
state. However, when the real part of G function is also positive
below the threshold, the pole can still be found in the model T
matrix with a repulsive potential. These poles far below threshold
are beyond the valid region of the model approximation and should be
discarded. Since varying the $G$ function in a reasonable range does
not influence our conclusion qualitatively, we present our numerical
results in the dimensional
regularization scheme with $a_\mu=-3.71$, 
in this paper.

\begin{table}[htbp]
 \renewcommand{\arraystretch}{1.2}
\centering \caption{ The parameters for two types of G functions
in the cases of $K\Sigma$, $\bar D\Sigma_c$ and $B\Sigma_b$
interactions, with $a_\mu$ for Eq.(\protect\ref{Gf}) and $\Lambda$
for Eq.(\protect\ref{Gf2}). The listed $a_\mu$ and $\Lambda(GeV)$
give the same value of two G functions at the corresponding
threshold.}\label{gfunt}\vspace{0.cm}
\begin{tabular}{l|cccccc}
                       & Threshold(GeV)   &  \multicolumn{5}{c}{$a_\mu$}\\
 \hline
   $\Lambda(GeV)$      &                  & 0.7         & 0.8         & 0.9       & 1.0        & 1.1                                  \\
   $B \Sigma_{b}$      & $11.087$         & $-3.679$    & $-3.715$    & $-3.751$  & $-3.786$   & $-3.822$                                        \\
   $\bar{D}\Sigma_c$   & $4.231$          & $-2.196$    & $-2.283$    & $-2.369$  & $-2.453$   & $-2.536$                                         \\
   $K\Sigma$           & $1.688$          & $-1.297$    & $-1.463$    & $-1.619$  & $-1.766$   & $-1.905$                                         \\
\end{tabular}
\end{table}

With the potential and $G$ function fixed, the unitary T amplitude
can be obtained by Eq.(\ref{Bethe}). The poles in the $T$ matrix are
looked for in the complex plane of $\sqrt{s}$. Those appearing in
the first Riemann sheet below threshold are considered as bound
states whereas those located in the second Riemann sheet and above
the threshold of some channel are identified as resonances. As
previously discussed, the poles will be kept only when the real part
of Eq.(\ref{Gf}) is negative.

From the T matrix for the $PB\to PB$ and $VB\to VB$ coupled-channel
systems, we can find the pole positions $z_R$. Six poles are found
in the real axes below corresponding thresholds and therefore they
are bound states. For these cases the coupling constants are
obtained from the amplitudes in the real axis. These amplitudes
behave close to the pole as:
\begin{eqnarray}
T_{ab}=\frac{g_{a}g_{b}}{\sqrt{s}-z_{R}} .
\end{eqnarray}
We can use the residue of $T_{aa}$ to determine the value of
$g_{a}$, except for a global phase. Then, the other couplings are
derived from
\begin{eqnarray}
g_{b}=\lim_{\sqrt{s}\rightarrow
z_{R}}(\frac{g_{a}T_{ab}}{T_{aa}})\ .\label{coupling2}
\end{eqnarray}

\section{Numerical results for the super-heavy $N^*$ and $\Lambda^*$ }
\label{s3}

Firstly, we discuss the (I, S) = (1/2, 0) sector. There are two
channels, $B\Sigma_b$ and $B\Lambda_b$. The masses of these
particles are taken from~\cite{PDG}, $m_{B}=5.279$~GeV,
$m_{B^*}=5.325$~GeV, $m_{\Sigma_b}=5.808$~GeV and
$m_{\Lambda_b}=5.620$~GeV. With the approach outlined in the last
section, the obtained pole positions $z_R$ and coupling constants
$g_\alpha$ are listed in Tables \ref{nos} for $PB \to PB$ and $VB
\to VB$. Because these poles are bound states for each channel, they
have zero width when neglecting transitions mediated by t-channel
exchange of heavy beauty mesons. To consider some possible decay
channels for them, such as $\pi N$, $\eta N$, $K\Sigma$, $\eta_b N$
and so on, we estimate these decays through heavy beauty meson
exchanges by means of box diagrams as in
Refs.\cite{raquel,geng,charm}. We neglect transitions to the hidden
charm channels such as $\bar{D}\Sigma_{c}$ and
$\bar{D}\Lambda^+_{c}$, because they need t-channel exchange of too
heavy vector meson constituted of charm and beauty quarks. The
results for $PB$ and corresponding $VB$ channels are listed in
Table~\ref{noswidth}. Comparing results in Table~\ref{nos} and
Table~\ref{noswidth}, the influence of these additional coupled
channels to the masses of predicted states is negligible. This is
because the transition potential by exchanging heavy beauty vector
meson is much smaller than the potential by exchanging light vector
meson.

\begin{table}[ht]
      \renewcommand{\arraystretch}{1.1}
     \setlength{\tabcolsep}{0.35cm}
\begin{center}\caption{Pole positions $z_R$ and coupling
constants $g_a$ for the states in (I, S) = (1/2, 0)
sector.}\label{nos}
\begin{tabular}{ccc}\hline
  $z_R$ (MeV)    & \multicolumn{2}{c}{$g_\alpha$}\\
\hline
                 &  $B \Sigma_{b}$ & $B \Lambda_{b}$ \\
          $11052$&  $2.05$               &  $0$\\
\hline
                 & $B^{*} \Sigma_{b}$ & $B^{*} \Lambda_{b}$ \\
          $11100$& $2.02$                   &  $0$\\
\hline\end{tabular}
\end{center}
\end{table}

\begin{table}[htbp]
     \setlength{\tabcolsep}{0.15cm}
\begin{center}\caption{Mass ($M$), total width ($\Gamma$), and
partial decay widths ($\Gamma_i$) for (I, S) = (1/2, 0)
sector.}\label{noswidth}
\begin{tabular}{ccccccccc}\hline
 $M$ (MeV)   & $\Gamma$ (MeV)  & \multicolumn{5}{c}{$\Gamma_i$ (MeV) }\\
\hline
       &             & $\pi N$   & $\eta N$   & $\eta' N$      & $K \Sigma$ & $\eta_bN$\\
$11052$& $1.38$      & $0.10$    & $0.21 $    & $0.11$         & $0.42$     & $0.52$\\
\hline
       &             & $\rho N$  & $\omega N$ & $K^{*} \Sigma$ &  $\Upsilon N$\\
$11100$& $1.33$      & $0.09$    & $0.30$     &  $0.39$        &  $0.51$    \\
\hline\end{tabular}
\end{center}
\end{table}

We also do not consider the coupled channel effect between $VB$ and
$PB$ channels as in Ref.\cite{charm}. The reason is that the
transition potentials $PB\to VB$ are much smaller than the
potentials of $PB\to PB$ or $VB\to VB$. Taking $B\Sigma_b \to
B^*\Sigma_b$ through t-channel pion exchange as an example, the $B^*
\pi B$ coupling is proportional to $(p_{B}-p_{\pi})_\mu
\varepsilon^\mu_{B^*}$ and is zero in the static limit which ignores
the three momenta of mesons and assumes
$\varepsilon^\mu_{B^*}=(0,\vec{\varepsilon}_{B^*})$. Going beyond
the static limit will give a non-zero transition potential but still
much smaller than its diagonal partners. This has been demonstrated
by the production rate of $J/\psi / \eta_c$ from $\bar pp$
collisions in Ref.\cite{charm}. The cross section for $\bar
pp\to\bar pp J/\psi$ through $\bar D\Sigma_c$ bound state is smaller
than that for $\bar pp\to\bar pp \eta_c$ by more than an order of
magnitude for similar excess energies. Therefore, the coupled
channel effect between $VB$ and $PB$ channels is expected to have
negligible influence on our predicted states.

One problem associated to the beauty sector should be addressed
here. As shown in Fig.\ref{pgf}, the loop functions of the hidden
beauty sector, calculated with the cut-off or with dimensional
regularization, show a quite different energy dependence and cannot
be made similar over a reasonable range of values, as is the case
for the hidden strange sector. This is due to that the on-shell
momentum in the beauty channel shows a much stronger energy
dependence than in the lightest channel. The results listed in
Tables~\ref{nos},\ref{noswidth} are obtained in the dimensional
regularization scheme, where the subtraction constant is adjusted to
the value of the $\Lambda=0.8$ GeV cut-off loop-function at
threshold. However, the binding energy is found to be about 35 MeV
for the $B\Sigma_b$ channel, which lies quite away of its threshold,
where the real parts of the two loop functions are very different.
This makes the choice of matching point for the two loop functions
questionable. In order to get some feeling about the choice of the
match point, it is also tried to match the two loop functions at 30
MeV below threshold. Then the regularization subtraction constant
moves from -3.715 to -3.774, and the binding energy moves from 35
MeV to 59 MeV. If the $\Lambda=0.8$ GeV cut-off loop function is
used directly, then the binding energy increases to 145 MeV. So the
simple Valencia model for the beauty sector works not as good as for
the hidden strange sector. The uncertainty for the concrete binding
energies is quite large of the order of tens to a hundred MeV. But
the qualitative conclusion for possible existence of bound state
should be very solid.

Then we discuss the (I, S) = (0, -1) sector. There are 3 channels,
$B_s\Lambda_b$, $B\Xi_b$ and $B\Xi'_b$. The masses of $B$, $B_s$,
$\Xi_b$ and $\Lambda_b$ have been precisely measured and can be
taken from Ref.\cite{PDG}. $m_{B_s}=5.366$~GeV,
$m_{B^*_s}=5.4128$~GeV and $m_{\Xi_b}=5.7924$~GeV. The $\Xi'_b$ has
not been observed yet. Its mass has been predicted to be 5.922 GeV
in Ref.\cite{wangzhigang} and 5.960 GeV in Ref.\cite{jenkins}. We
choose a middle value 5.940 GeV in this paper. From
Table~\ref{mcof}, the $B\Xi'_b$ channel is decoupled from other two
channels, so there should be a bound state for this channel, the
same as corresponding vector-meson-baryon channel, $B^*\Xi'_b$. For
this channel, the results are listed in Table~\ref{mspoles}.

For the coupled $B_s\Lambda_b$ and $B\Xi_b$ channels, the T matrix
can be written as:
\begin{equation}
T=\frac{1}{1-V'G_{B\Xi_b}}\left( \begin{array}{cc}
V^2_{B_s\Lambda_b \to B\Xi_b}G_{B\Xi_b} & V_{B_s\Lambda_b \to B\Xi_b} \\
V_{B_s\Lambda_b \to B\Xi_b} & V' \end{array} \right) \label{tmatrix}
\end{equation}
with $ V'=V_{B\Xi_b \to B\Xi_b}+V^2_{B_s\Lambda_b \to
B\Xi_b}G_{B_s\Lambda_b}$.

The $V'$ is negative and hence provides an attractive potential. For
$a_\mu=-3.71$, one pole is found for the coupled-channel system,
with mass between the two thresholds of $B_s\Lambda_b$ (10.986 GeV)
and $B\Xi_b$ (11.071 GeV). The pole position depends on the value of
$a_\mu$ as demonstrated in Table~\ref{mspoles} and can move to below
the $B_s\Lambda_b$ threshold when the magnitude of $a_\mu$
increases, such as for $a_\mu=-3.82$ corresponding to the
$\Lambda=1.1$~GeV.

\begin{table}[ht]
      \renewcommand{\arraystretch}{1.1}
     \setlength{\tabcolsep}{0.35cm}
\begin{center}\caption{Pole positions $z_R$ with different
$a_\mu$ for $PB \to PB$ in (I, S) = (0, -1) sector.}
\begin{tabular}{cccc}
\hline
 $\Lambda$(GeV)  & $a_\mu$          & \multicolumn{2}{c}{$z_R$ (MeV)} \\
\hline
                                & $B_{s} \Lambda_{b}$ and $B \Xi_{b}$   &  $B\Xi'_{b}$          \\
  $0.7$          &$-3.68$          &  $11030-0.60i$                        &  $11198$            \\
  $0.8$          &$-3.71$          &  $11021-0.59i$                        &  $11191$            \\
  $0.9$          &$-3.75$          &  $11004-0.49i$                        &  $11178$            \\
  $1.0$          &$-3.78$          &  $10990-0.24i$                        &  $11167$            \\
  $1.1$          &$-3.82$          &  $10970$                              &  $11151$            \\
\hline\end{tabular} \label{mspoles}
\end{center}
\end{table}

The coupling constants and the possible decay channels of these two
resonances are listed in Tables~\ref{msg} and \ref{mswidth} for
$a_\mu=-3.71$. Similarly, the results for the corresponding
vector-meson-baryon channels are also listed in Tables~\ref{msg} and
\ref{mswidth} for $a_\mu=-3.71$.

\begin{table}[ht]
      \renewcommand{\arraystretch}{1.1}
     \setlength{\tabcolsep}{0.35cm}
\begin{center}\caption{Pole positions $z_R$ and coupling
constants $g_a$ for the states in (I, S) = (0, -1) sector for
$a_\mu=-3.71$.}
\begin{tabular}{cccc}\hline
  $z_R$ (MeV)    & \multicolumn{3}{c}{$g_\alpha$}\\
\hline
                 & $B_{s} \Lambda_{b}$   & $B \Xi_{b}$        & $B \Xi'_{b}$\\
  $11021-0.59i$  & $0.14-0.11i$          & $2.27+0.004i$      & $0$              \\
  $11191$        & $0$                   & $0$                & $1.92$           \\
\hline
                 & $B^*_{s} \Lambda_{b}$   & $B^* \Xi_{b}$    & $B^* \Xi'_{b}$\\
  $11069-0.59i$  & $0.14-0.12i$            & $2.24+0.005i$    & $0$              \\
  $11238$        & $0$                     & $0$              & $1.89$              \\
\hline\end{tabular} \label{msg}
\end{center}
\end{table}

\begin{table}[ht]
     \setlength{\tabcolsep}{0.15cm}
\begin{center}\caption{Mass ($M$), total width ($\Gamma$), and
partial decay widths ($\Gamma_i$) for the states in (I, S) = (0, -1)
sector for $a_\mu=-3.71$.}
\begin{tabular}{ccccccccc}\hline
 $M$ (MeV)  & $\Gamma$ (MeV) & \multicolumn{6}{c}{$\Gamma_i$} (MeV)\\
\hline
       &          & $\bar{K} N$  & $\pi\Sigma$ & $\eta\Lambda$ & $\eta'\Lambda$ & $K\Xi$   & $\eta_b\Lambda$  &$B_s\Lambda_b$\\
$11021$& $2.21$   & $0.65$       & $0.01$      & $0.08 $        & $0.14$        & $0.01$   & $0.19$           & $1.18$  \\
$11191$& $1.24$   & $0 $         & $0.28$      & $0.18 $        & $0.10$        & $0.18 $  & $0.48$           & $0$\\
\hline
       &         & $\bar K^*N$ & $\rho\Sigma$ & $\omega\Lambda$ & $\phi\Lambda$ & $K^*\Xi$ & $\Upsilon\Lambda$ &$B^*_s\Lambda_b$\\
$11070$& $2.17$  & $0.61$      & $0.01$       & $0.01$          & $0.20$        & $0.01$   & $0.19$            & $1.18$\\
$11239$& $1.19$  & $0$         & $0.26$       & $0.26$          & $0$           & $0.17$   & $0.48$            & $0$\\
\hline\end{tabular}
 \label{mswidth}
\end{center}
\end{table}

Totally two $N^*$ and four $\Lambda^*$ states are predicted to exist
with masses above $11$ GeV and very narrow widths of only a few MeV.
The very narrow widths are due to the fact that all decays are tied
to the necessity of the exchange of a heavy beauty vector meson
because of hidden $b\bar{b}$ components involved in these states,
and hence are suppressed. If these predicted narrow $N^*$ and
$\Lambda^*$ resonances with hidden beauty would be found, they
definitely cannot be accommodated by quark models with three
constituent quarks. Together with other possible $N^*$ and
$\Lambda^*$ states of other quantum numbers with hidden beauty, they
should form a super-heavy island of the heaviest masses for excited
nucleons $N^*$ and excited hyperons $\Lambda^*$.

\section{Effects of momentum dependent terms in the potential}
\label{s4}

For our model calculations in the last two sections, the static
limit is assumed for the t-channel exchange of light vector mesons
by neglecting momentum dependent terms as discussed after the
Eq.(\ref{eq:lag}). However, in Ref.\cite{jrv}, dynamically generated
open charmed baryons were studied by solving the Lippmann -
Schwinger equation beyond the zero range approximation. The momentum
dependent terms were found to have non-negligible effects on the
results. In order to investigate the possible influence of the
momentum dependent terms in this case, in this section, we use the
conventional Schrodinger Equation approach to study possible bound
states for the $B\Sigma_b$ channel by keeping the momentum dependent
terms in the t-channel meson exchange potential.

The deduction of the momentum dependent potential by the t-channel
exchange of light vector mesons is straightforward. By keeping
momentum dependent terms up to quadratic order with proper
normalization factor for the Schrodinger Equation and including the
vertex form factors as in the Bonn potential model~\cite{Bonn}, the
effective S-wave $B\Sigma_b$ potential is obtained as the following:
\begin{eqnarray}
V^{S}_{ab(P_{1}B_{1}\to
P_{2}B_{2})}&=&\frac{C_{ab}m^2_{V}}{4f^2}\frac{1}{\vec{q}^{2}+m_V^2}
\left(\frac{\Lambda^2_V-m_V^2}{{\Lambda}^2_V+\vec{q}^2}\right)^2\times\nonumber\\
&&\left(1 + \frac{m_P^2 + 2m_B^2 + 4 m_P m_B}{4 m_P^2
m_B^2}\vec{k}^{2} + \frac{2 m_B^2 - m_P^2}{16 m_P^2
m_B^2}\vec{q}^{2}\right),\label{vsqq}
\end{eqnarray}
where $\vec{k}$ and $\vec{q}$ are defined as $(\vec{p}+\vec{p'})/2$
and $\vec{p}-\vec{p'}$ with $\vec{p}$ and $\vec{p'}$ the initial and
final momenta of the pseudo-scalar meson, respectively, in the
center of mass system of the $B\Sigma_b$ channel. For simplicity, we
assume the same cut-off parameter $\Lambda_V$ for the $\rho$ and
$\omega$ mesons.

The effective potential for the Schrodinger Equation in the
coordinate space, $V(\vec r)$, can be obtained by using the
following Fourier-transformation formulae:
\begin{eqnarray}
   \mathcal{F}\{ {(\frac{\Lambda^2-m^2}{{\Lambda}^2+\vec{q}^2})}^2\frac{1}{\vec{q}^2+m^2}\}&=&
 \frac{1}{4\pi} \left(\frac{e^{-mr}}{r}-\frac{e^{-\Lambda r}}{r}-(\Lambda^2-m^2)\frac{e^{-\Lambda r}}{2 \Lambda}\right),\nonumber\\
   \mathcal{F}\{ {(\frac{\Lambda^2-m^2}{{\Lambda}^2+\vec{q}^2})}^2{\frac{\vec{q}^2}{\vec{q}^2+m^2}}\}&=&
 \frac{1}{4\pi} \left(m^2(-\frac{e^{-mr}}{r}+\frac{e^{-\Lambda r}}{r}) +(\Lambda^2-m^2)\frac{\Lambda e^{-\Lambda r}}{2} \right),\nonumber\\
   \mathcal{F}\{{(\frac{\Lambda^2-m^2}{{\Lambda}^2+\vec{q}^2})}^2{\frac{\vec{k}^2}{\vec{q}^2+m^2}}\}&=&
 \frac{1}{4\pi}\left(\frac{m^2}{4}\frac{e^{-mr}}{r}-\frac{\Lambda ^2}{4}\frac{e^{-\Lambda r}}{r}-\frac{\Lambda^2-m^2}{4}(\frac{\Lambda r}{2}-1)
 \frac{e^{-\Lambda r}}{r} \right) \nonumber\\
   &&-\frac{1}{8\pi}\{\nabla^2,\frac{e^{-mr}}{r}-\frac{e^{-\Lambda r}}{r}-\frac{\Lambda^2-m^2}{2}
 \frac{e^{-\Lambda r}}{\Lambda}\} .\nonumber
\end{eqnarray}

Then we can solve the Schrodinger Equation
\begin{equation}
(-\frac{\hbar^2}{2\mu}\nabla^{2}+V(\vec{r})-E)\Psi(\vec{r})=0 ,
\end{equation}
to find possible bound state with eigenvalue E and corresponding
wave function $\Psi(\vec r)$, and estimate the size of the system
$\bar r$ with the formula
\begin{eqnarray}
\bar{r}&=&\sqrt{\int r^2drd\Omega
\Psi^*(\vec{r})r^2\Psi(\vec{r})\label{r}}.
\end{eqnarray}

It is found that whether there exists a bound state depends on the
cut-off parameter $\Lambda_V$. The results corresponding to various
$\Lambda_V$ values are listed in Table.\ref{eigenvalues}.

\begin{table}
        \renewcommand{\arraystretch}{1.1}
         \setlength{\tabcolsep}{0.15cm}
\begin{center}\caption{Eigenvalue $E$ and average size of
system $\bar r$ vs the cut-off parameter $\Lambda_V$.}
\begin{tabular}{ccccccccccccc}\hline
          & $\Lambda_V$(MeV)       & \ 1100  & 1200  & 1300  & 1400  & 1500   & 1600   & 1700   & 1800   & 1900    & 2000\\
          & $E$(MeV)             & \  -    & -0.85 & -4.49 & -10.5 & -18.4  & -27.9  & -38.7  & -50.5  & -63.3   & -78.9\\
          & $\bar{r}$(fm)        & \  -    & 2.36  & 1.19  & 0.86  & 0.70   & 0.60   & 0.53   & 0.48   & 0.44    & 0.41\\
\hline\end{tabular} \label{eigenvalues}
\end{center}
\end{table}

From the Table~\ref{eigenvalues}, we can see that when the cut-off
parameter $\Lambda_V$ is $1200$ MeV or larger, the effective
potential can provide enough attraction to form a bound state. For
$\Lambda_V$ in the range of $1200\sim 1800$ MeV, the binding energy
is in the range of $1\sim 50$ MeV with the average distance between
two hadrons to be about $0.5\sim 2$ fm. The typical values for
$\Lambda_V$ in the Bonn potential are $\Lambda_\rho=1400$ MeV and
$\Lambda_\omega=1500$ MeV~\cite{Bonn}. The binding energy
corresponding to $\Lambda_V=1600$ MeV is quite close to that
obtained by the Valencia approach in the last section. This gives
some justification of the simple Valencia approach although there
could be an uncertainty of $10\sim 20$ MeV for the binding energy.

According to Ref.\cite{Ericson}, ``the apparent radius of the pion
as seen by the photon is determined almost completely by the
intermediate $\rho$ meson: the intrinsic pion size must be
considerably smaller than the measured charge radius. In
descriptions which explicitly include the $\rho$ meson, the pion can
therefore be considered point-like for all practical purposes". In
our approach with the t-channel $\rho$ meson exchange explicitly
included, the D meson similar to the pion is expected to have very
small size while the intrinsic radius of $\Sigma_c$ baryon is
expected to be around 0.5 fm similar to that for the
proton~\cite{Ericson}. With typical size $\bar r$ larger than 0.5
fm, our predicted hadron molecular state should not suffer much from
internal structure of the constituents.

For the $B_s\Lambda_b$-$B\Xi_b$ coupled channel case, it is not so
easy to use the Schrodinger Equation approach. Since the simple
Valencia approach gives a consistent result for the $B\Sigma_b$
single channel case with Schrodinger Equation approach, we expect it
also gives reasonable results for the $B_s\Lambda_b$-$B\Xi_b$
coupled channel case.

\section{Production of $N^*_{b\bar b}$ and $\Lambda^*_{b\bar b}$ in $pp$ and $ep$ collisions }
\label{s5}

\begin{figure}[htbp] \vspace{-0.cm}
\begin{center}
\includegraphics[width=0.35\columnwidth]{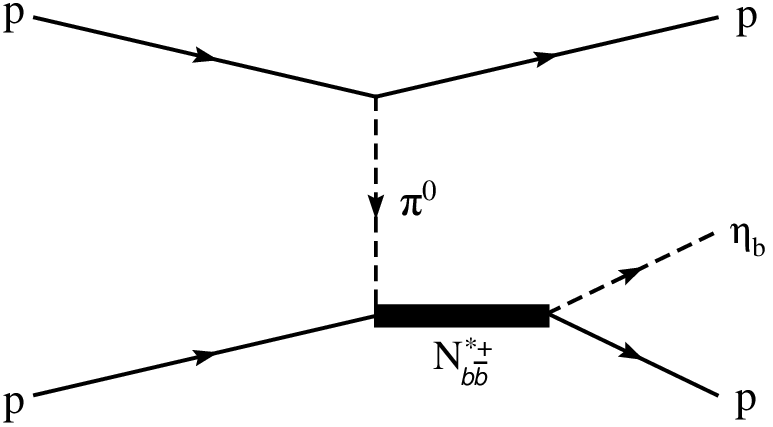}
\includegraphics[width=0.35\columnwidth]{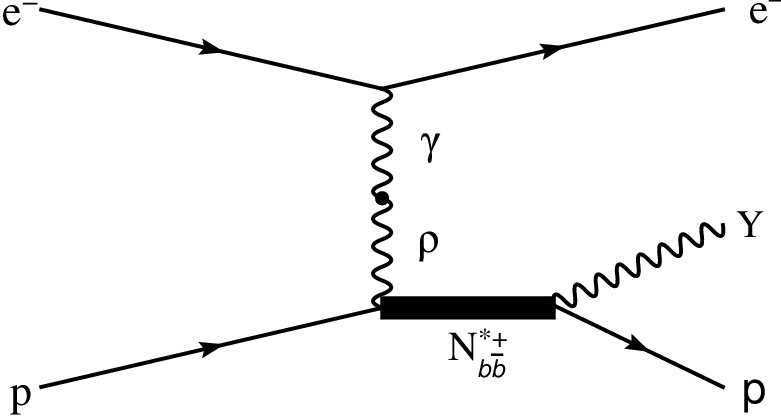}
\caption{Feynman diagrams for the reaction $pp \to pp\eta_b$ and $ep
\to ep\Upsilon$.} \label{feexp}
\end{center}
\end{figure}

In order to look for these predicted super-heavy $N^*_{b\bar b}$ and
$\Lambda^*_{b\bar b}$ states, we give an estimation of their
production cross sections in the $pp \to pp\eta_b$ and $ep \to
ep\Upsilon$ reactions. The Feynman diagrams are shown in
Fig.\ref{feexp}. We also estimate the background of the $pp \to
pp\eta_b$ with $N^*_{b\bar{b}}$ replaced by the nucleon pole.

The Lagrangians for the interaction vertices of these two reactions
are as follows~\cite{xie,zouf,wujj}:
\begin{eqnarray}
{\cal L}_{NN\pi}&=&g_{NN\pi}\bar{N}\gamma_5\vec{\tau}\cdot\vec{\psi}_{\pi}N+h.c.,\\
{\cal L}_{NN\eta_b}&=&g_{NN\eta_b}\bar{N}\gamma_5\psi_{\eta_b}N+h.c.,\\
{\cal L}_{N^{*}_{b\bar{b}}N\pi}&=&g_{N^{*+}_{b\bar{b}}N\pi}\overline{N^{*}_{b\bar{b}}}N\vec{\tau}\cdot\vec{\psi}_{\pi}+h.c.,\\
{\cal L}_{N^{*}_{b\bar{b}}N\eta_b}&=&g_{N^{*+}_{b\bar{b}}N\eta_b}\overline{N^{*}_{b\bar{b}}}N\psi_{\eta_b}+h.c.,\\
{\cal L}_{ee\gamma}&=&ie\bar{\psi}_{e}\gamma_{\mu}\psi_{e}A^{\mu}_{\gamma}+h.c.,\\
{\cal L}_{\rho\gamma}&=&\frac{em^{2}_{\rho}}{f_{\rho}}\rho^{\mu}A_{\gamma\mu}+h.c.,\\
{\cal L}_{N^{*}_{b\bar{b}}N\rho}&=&g_{N^{*}_{b\bar{b}}N\rho}\overline{N^{*}_{b\bar{b}}}\gamma_{5}\gamma^{\mu}N\tilde{g}_{\mu\nu}(P_{N^{*}_{c\bar{c}}})\vec{\tau}\cdot\vec{\psi}^{\nu}_{\rho}+h.c.,\\
 {\cal L}_{N^{*}_{b\bar{b}}N\Upsilon}&=&g_{N^{*}_{b\bar{b}}N\rho}\overline{N^{*}_{b\bar{b}}}\gamma_{5}\gamma^{\mu}N\tilde{g}_{\mu\nu}(P_{N^{*}_{c\bar{c}}})\psi^{\nu}_{\Upsilon}+h.c..
\end{eqnarray}
with
$\tilde{g}_{\mu\nu}(P)=-g_{\mu\nu}+\frac{P^{\mu}P^{\nu}}{P^{2}}$.

In our model calculation, we only consider S-wave PB and VB
interactions, so the spin-parity $J^{P}$ of our predicted
$N^*_{b\bar{b}}$ for the PB channels is $1/2^{-}$, and the
$N^*_{b\bar{b}}$ for the VB channels can be either $1/2^{-}$ or
$3/2^{-}$, but assumed to be $1/2^{-}$ here for a simple estimation
of rough production rate. The coupling constants of the Lagrangians
can be either calculated from its corresponding partial decay widths
or obtained from references. They are all listed in
Table~\ref{coupling}. For the $NN\eta_b$ vertex, the width of
$\eta_b$ has not been measured. Since both $\eta_b$ and $\eta_c$
couple to nucleon through two gluon exchange, we use the relation
$g_{NN\eta_b}\sim
g_{NN\eta_c}\alpha_s^4(M_{\eta_b})/\alpha_s^4(M_{\eta_c})$ to
estimate the $g_{NN\eta_b}$ with $g_{NN\eta_c}$ determined from the
decay width of $\eta_c\to p\bar p$.

\begin{table}[ht]
     \setlength{\tabcolsep}{0.15cm}
\begin{center}\caption{The coupling constants of involved
vertices and corresponding widths used.}
\begin{tabular}{ccccccccc}\hline
 Vertex                            & $\Gamma(MeV)$       & Coupling Constant($g^2/4\pi$)\\
\hline
 $pp\pi^0$                         &                     & $14.4$ \\
 $N^{*+}_{b\bar{b}}p\pi^0$         & $0.033$             & $1.03\times 10^{-5}$ \\
 $N^{*+}_{b\bar{b}}p\eta_b$        & $0.52$              & $1.81\times 10^{-3}$ \\
 $ee\gamma$                        &                     & $1/137$ \\
 $\gamma\rho$                      &                     & $2.7$~\protect\cite{xie} \\
 $N^{*+}_{b\bar{b}}p\rho^0$        & $0.030$             & $1.55\times 10^{-8}$ \\
 $N^{*+}_{b\bar{b}}p\Upsilon$      & $0.51$              & $4.72\times 10^{-4}$ \\
 $pp\eta_{b}$                      &                     & $1\times 10^{-6}$ \\
\hline\end{tabular}  \label{coupling}
\end{center}
\end{table}

As usual, the off-shell form factors should be considered here. We
use two kinds of form factors for mesons and baryons, respectively.
\begin{eqnarray}
F_{M}&=&\frac{\Lambda_{M}^{2}-m^{2}_{M}}{\Lambda^{2}_{M}-p^{2}_{M}},\\
F_{N}&=&\frac{\Lambda_{N}^{4}}{\Lambda_{N}^{4}+(p^{2}_{N}-m^{2}_{N})^{2}},
\end{eqnarray}
where $M$ stands for $\pi$ or $\rho$, and $N$ stands for
$N^{*}_{b\bar{b}}$ or nucleon pole. Here $\Lambda_{M}=1.3$~GeV,
$\Lambda_{N}=1.0$~GeV.

\begin{figure}[htbp] \vspace{-0.cm}
\begin{center}
\includegraphics[width=0.49\columnwidth]{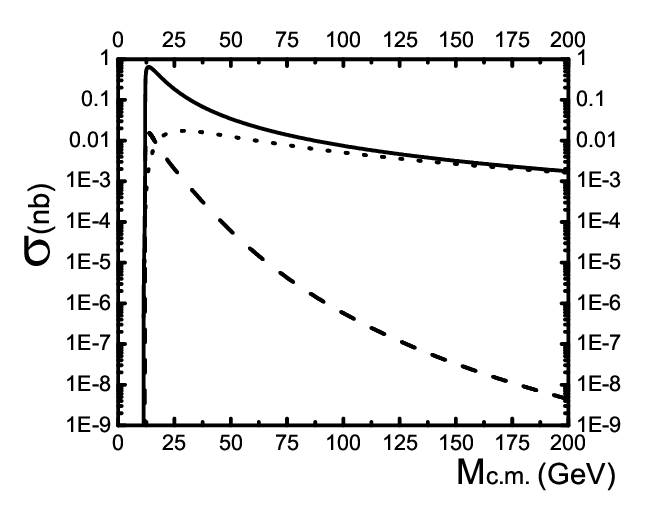}
\includegraphics[width=0.49\columnwidth]{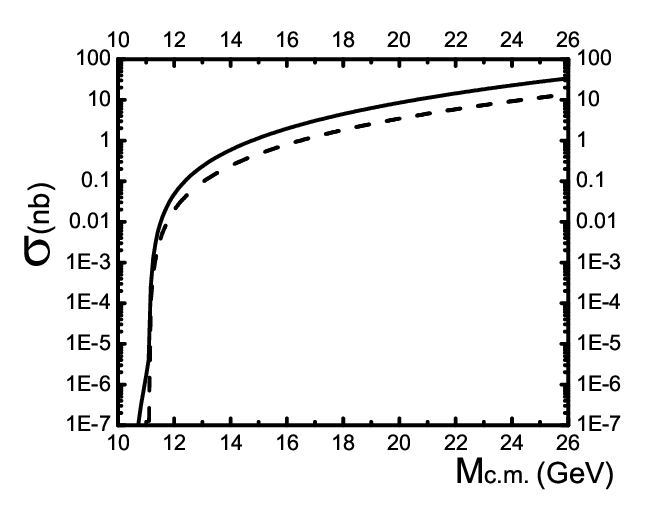}
\caption{Total cross section vs invariant mass of system for $pp \to
pp \eta_b$ reaction (left) and $e^-p \to e^-p \Upsilon$ reaction
(right), with (dashed curves) and without (solid curves) including
off-shell form factors, through production of the predicted
$N^*_{b\bar b}$ resonances. The dotted curve is the background
contribution from the nucleon pole for the $pp \to pp \eta_b$
reaction without including form factors.} \label{exp}
\end{center}
\end{figure}

To produce the predicted $N^*_{b\bar{b}}(11052)$ in the pp
collisions, the center-of-mass energy should be above 12 GeV. In
Fig.\ref{exp}, the left figure shows our theoretical estimated total
cross section for the $pp \to pp \eta_b$ reaction through the
$N^*_{b\bar{b}}$ production vs the center-of-mass energy, with
(dashed curve) and without (solid curve) including the off-shell
form factors. As an estimation of background contribution to the
$N^*_{b\bar{b}}$ production, we also calculate the corresponding
cross section through the off-shell nucleon pole without including
the form factors. The result is shown by the dotted curve. The
contribution from the nucleon pole is much smaller than that from
the $N^*_{b\bar{b}}$ production, because the nucleon pole is much
more off-shell than $N^*_{b\bar{b}}$. The contribution of the
nucleon pole with form factors becomes very small for the same
reason, so it is not shown in Fig.\ref{exp}. This background
reaction will not influence the observation of the $N^*_{b\bar{b}}$
production, especially for the energy range of $13\sim 25$~GeV. The
cross section from $N^*_{b\bar{b}}$ production is about 0.1 nb,
which is much smaller than that for the corresponding reaction $pp
\to pp \eta_c$ with $N^*_{c\bar{c}}$ production~\cite{charm} of
about 0.1 $\mu b$. The main reason is that both couplings of
$N^*_{b\bar{b}}N\pi$ and $N^*_{b\bar{b}}N\eta_b$ are much smaller
than the corresponding $N^*_{c\bar{c}}N\pi$ and
$N^*_{c\bar{c}}N\eta_c$ couplings. These two vertices cause a
reduction of about 2 orders of magnitude. In addition, because the
center-of-mass energy here is much larger than that in the previous
calculation for the $\eta_c$ production, the propagator of exchanged
$\pi^0$ further reduces the contribution. For the same reason, the
contribution with form factors is much less than that without them.

For the production of $N^*_{b\bar{b}}(11100)$ in $ep$ collisions,
the invariant mass of the system should be above 11 GeV. The right
figure in Fig.\ref{exp} shows our calculated total cross section for
the $e^-p \to e^-p \Upsilon$ reaction vs the invariant mass of the
system with (dashed curve) and without (solid curves) including form
factors. The cross section of this reaction is much larger than that
for the $pp \to pp \eta_b$ reaction. The reason is due to the
propagator of massless photon. The propagator of photon is given as
the following:
\begin{eqnarray}
\frac{1}{p^{2}_{\gamma}}&=&\frac{1}{2(m^2_{e}+p_{i}p_{f}cos\theta-E_{i}E_{f})},\label{phog}
\end{eqnarray}
where the $p_{i}$, $E_{i}$ are the three-momentum and energy of
initial $e^-$, and $p_{f}$, $E_{f}$ for final $e^-$. $\theta$ is the
angle between initial and final $e^-$. When the directions of
initial and final $e^-$ are the same, {\sl i.e.}, $cos\theta=1$, the
value of Eq.(\ref{phog}) becomes very large because of the very
small mass of $e^-$. As the beam momentum of $e^-$ becomes larger,
the propagator of photon can reach very big value. For the invariant
mass of the system less than 15~GeV, the cross section of $e^-p \to
e^-p \Upsilon$ reaction is of the same order of magnitude as that of
$pp \to pp \eta_b$ reaction.

\section{Summary}
\label{s6}

In summary, the meson-baryon coupled channel unitary approach with
the local hidden gauge formalism is extended to the hidden beauty
sector. Two $N^*_{b\bar b}$ states and four $\Lambda^*_{b\bar b}$
states are predicted to be dynamically generated from coupled PB and
VB channels with the same approach as for the hidden charm
sector~\cite{charm}. Because of the hidden $b\bar{b}$ components
involved in these states, the masses of these states are all above
11 GeV while their widths are of only a few MeV, which should form
part of the heaviest island for the quite stable $N^*$ and
$\Lambda^*$ baryons. The nature of these states is similar as
corresponding $N^*_{c\bar c}$ and $\Lambda^*_{c\bar c}$ states
predicted in Ref.\cite{charm}, which definitely cannot be
accommodated by the conventional 3q quark models.

Production cross sections of the predicted $N^*_{b\bar{b}}$
resonances in $pp$ and $ep$ collisions are estimated as a guide for
the possible experimental search at relevant facilities in the
future. For the $pp \to pp \eta_b$ reaction, the best center-of-mass
energy for observing the predicted $N^*_{b\bar{b}}$ is $13\sim 25$
GeV, where the production cross section is about 0.01 nb. For the
$e^-p \to e^-p \Upsilon$ reaction, when the center-of-mass energy is
larger than 14 GeV, the production cross section should be larger
than 0.1 nb. Nowadays, the luminosity for pp or ep collisions can
reach $10^{33}cm^{-2}s^{-1}$, this will produce more than 1000
events per day for the $N^*_{b\bar{b}}$ production. We expect future
facilities, such as proposed electron-ion collider (EIC)~\cite{EIC},
to discover these very interesting super-heavy $N^*$ and $\Lambda^*$
with hidden beauty.

\section*{Acknowledgments}
This work is supported by the National Natural Science Foundation of
China (NSFC) under grants Nos. 10875133, 10821063, 11035006 and by
the Chinese Academy of Sciences under project No. KJCX2-EW-N01, and
by the Ministry of Science and Technology of China (2009CB825200).

\end{document}